\documentclass[aps,prl,reprint]{revtex4-2}

\usepackage{graphicx}
\usepackage{dcolumn}
\usepackage{bm}

\begin{document}

\title{\textbf{Stretched-Exponential Aging Governs Nonequilibrium Precipitate Patterns} 
}%

\author{Amari Z. Morris}
\author{Oliver Steinbock}
 \email{Contact author: osteinbock@fsu.edu}
\affiliation{%
Florida State University, Department of Chemistry and Biochemistry, Tallahassee, FL 32306-4390, USA}%

\date{\today} 

\begin{abstract}
Localized growth in driven materials is often governed by intermittent failure, yet how a material's history biases failure sites remains poorly understood. Using pause-restart experiments on chemical precipitate membranes, we quantify the probability of age-dependent breaching. We show that the kinetics follow a stretched-exponential aging law with parameters that obey one-parameter scaling. As the system approaches a critical concentration, the stretching exponent $\beta$ tends to zero, signaling a crossover to scale-free, power-law behavior. A stochastic cellular automaton based on this aging rule reproduces the emergent filaments and their concentration-dependent thickening. Our findings identify aging-controlled failure with long-lived but decaying memory as a general route to pattern formation in far-from-equilibrium systems.
\end{abstract}


\maketitle


Materials that exist only under nonequilibrium conditions represent a largely unexplored frontier of condensed and soft matter physics \cite{Sarkar2025,Li2022,Shklyaev2024,Marchetti2013}. In such systems, continual energy input can in principle maintain structures that would otherwise disintegrate, enabling dynamic reconfiguration, repair, and growth across length scales \cite{Knoll2024,Zhu2024}. These behaviors require localized failure events whose probability depends on a material’s history, linking far-from-equilibrium matter to broader concepts of aging, memory, and fracture familiar from glasses, soft solids, and dissipative media \cite{Keim2014,Paulsen2014,Cipelletti2005}. Yet, with few exceptions, such adaptive dynamics remain rare in inorganic materials and are largely confined to living systems, such as growing plants and remodelling tissues \cite{AguilarHidalgo2018,Trepat2009,Dervaux2008}.

A chemical system that approaches these dynamics is the class of inorganic precipitation structures known as chemical gardens \cite{Barge2015,Nakouzi2016,Cardoso2020,Pantaleone2009}. These structures form far from equilibrium when a metal salt crystal is introduced into silicate or related solutions, producing self-organized hollow tubes with micrometer- to millimeter-scale diameters \cite{Cartwright2002,Guler2023}. The tube walls consist of thin, insoluble precipitates, typically metal hydroxides coated with silica. Growth is driven by osmotic pressure differences between the inner and outer solutions that sustain fluid flow and continuous dissolution of the salt seed \cite{Rauscher2018,Zheng2026,Bohner2015}. In closed structures, expansion proceeds either through stretching of compliant wall segments or through local breaches that allow the interior solution to surge outward \cite{Zheng2025}.

More controlled versions of the chemical garden experiment replace the seed crystal with a salt solution that is steadily injected into the precipitation-inducing reaction partner \cite{ThouvenelRomans2003,Batista2014,Kubodera2025}. Such studies identified distinct growth dynamics and morphologies and allowed for quantitative analysis of the tube diameter \cite{ThouvenelRomans2004}. Further simplifications of the experiment reduce the spatial dimensionality by confinement to microfluidic channels or the quasi-two-dimensional gap between two parallel plates \cite{Nogueira2023,Haudin2014}. In Hele-Shaw cells with a central injection port, chemical gardens can take a variety of different macroscopic shapes, including expanding rings, irregularly layered membranes, and filament-like conduits that are close analogs of the three-dimensional precipitate tubes \cite{Haudin2014,Wagatsuma2017,Rocha2021,Rocha2022,Rieder2022,Facchini2025}.

A recent model suggested that pattern selection in these systems is controlled by the temporal aging of the precipitate membrane, but this hypothesis has not been tested experimentally and the underlying failure statistics remain unknown \cite{Batista2023}. Here, we directly test this hypothesis using pause-restart experiments that quantify the probability of age-dependent breaching. We find stretched exponential behavior, with tightly coupled parameters that cross over to scale-free failure kinetics near a critical concentration. A stochastic cellular automaton incorporating this rule reproduces the observed filament formation and concentration-dependent thickening. These results establish aging-controlled failure as a physical mechanism for pattern selection and identify chemical gardens as a minimal model for adaptive growth in far-from-equilibrium materials.\\


In our experiments, we slowly deliver $\mathrm{CoCl_2}$ solution from a central injection port into a thin layer of sodium silicate solution confined to a horizontal Hele-Shaw cell. In response, precipitates form nearly instantaneously  \cite{Zahoran2019} along the interface of the two liquids. These solid reaction products create a thin membrane that substantially slows further mixing and effectively compartmentalizes the reaction system into an inner and outer domain. However, as the injected liquid is incompressible, the membrane must continuously yield to accommodate the volume of the steadily arriving $\mathrm{CoCl_2}$ solution.

\begin{figure}[t]
\centering
\includegraphics[width=0.9\columnwidth]{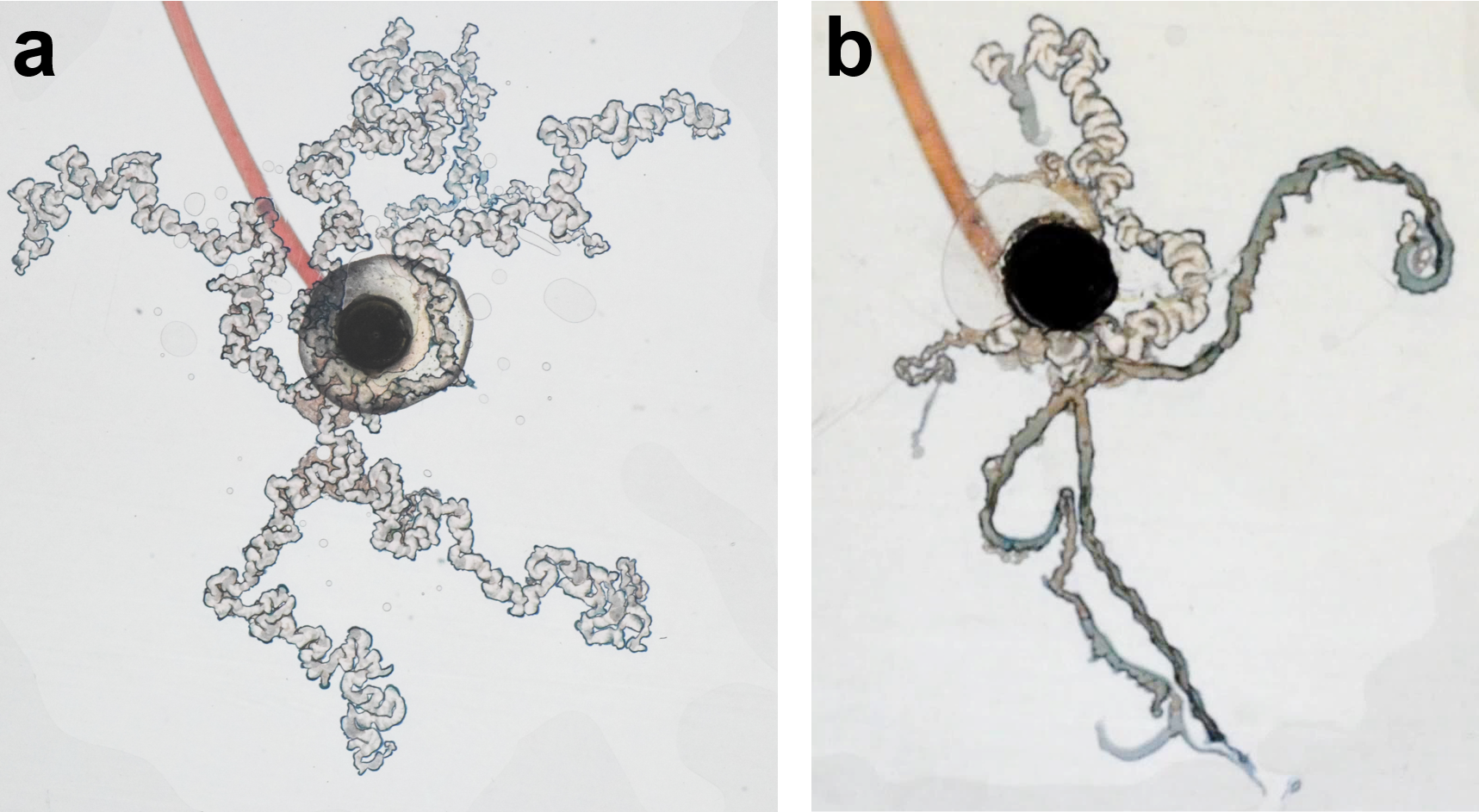}
\caption{Photos of precipitate patterns formed when pink $\mathrm{CoCl_2}$ solution is injected into a Hele-Shaw cell filled with clear sodium silicate solution. The black disks are injection ports connected to tubing. Initial concentrations: [$\mathrm{CoCl_2}$] = 0.625~M and (a,b) [silicate] = 4.0, 5.0~M, respectively. Image heights: 10.0 cm. Injection rate: (a) 1.15~mL/min and (b) 0.15~mL/min.}
\label{fig:images}
\end{figure}

Figure~\ref{fig:images} shows representative photos of patterns obtained at two concentrations $c$ of the optically transparent silicate solution and after 95~s of injection. The patterns consist of differently colored regions that have been studied spectroscopically \cite{Wang2017,Cartwright2011}. The pink interior is primarily due to Co(II) solution (a hexaaqua complex) and solid $\beta\mathrm{-Co(OH)_2}$. The turquoise blue color is caused by $\mathrm{[Co(OH)_4]^{2-}}$ ions in the hydrated, gel-like membrane. Darker features likely indicate further oxidation to $\mathrm{CoO(OH)}$ (brown) and $\mathrm{Co_3O_4}$ (black).

In agreement with earlier studies by Haudin et al. \cite{Haudin2014,Haudin2015,Haudin2015b}, we find that low silicate concentrations tend to produce fairly compact patterns that, upon closer inspection, reveal a labyrinthine structure (not shown). This structure arises from thin filaments at the pattern's perimeter that serve as the active growth zones. At higher silicate concentrations such as in Fig.~\ref{fig:images}, these filaments extend away from the injection site and form more clearly identifiable conduits. Filament width increases with silicate concentration. In addition, freshly formed segments preferentially adhere to nearby regions, producing a serpentine appearance.

All precipitate patterns in our experiments (and most related studies) grow only at a small number of distinct growth zones. These growth points are typically the tips of the finger-like structures. However, stochastic transitions can cause new filaments to emerge from older precipitates or active filaments to cease growth or branch. In addition, some experimental conditions can give rise to more irregular patterns in which active zones are short-lived and frequently switch location.

\begin{figure}[b]
\centering
\includegraphics[width=1.0\columnwidth]{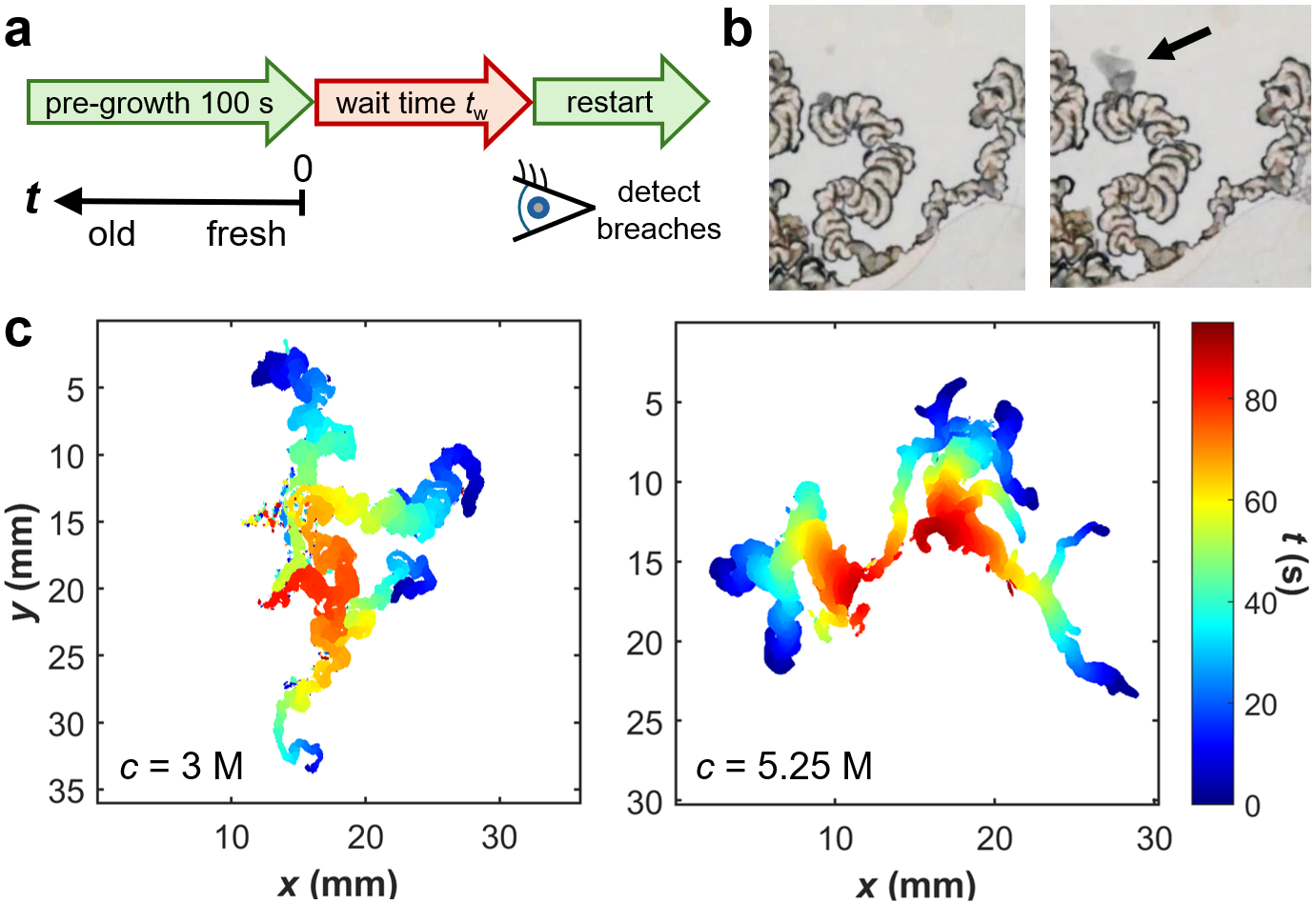}
\caption{
(a) Measurement protocol: the pattern is pre-grown, injection is paused for a wait time $t_w$, and growth is restarted while breach sites are optically detected. 
(b) Example of a paused pattern (left) and subsequent regrowth at the breach site (arrow). 
(c) Color-coded precipitate age maps for representative patterns at silicate concentrations $c=3$~M and 5.25~M; blue and red indicate fresh and old material, respectively.}
\label{fig:fig2}
\end{figure}

Because filament growth zones persist for extended times, we employ a pause–restart protocol to increase the likelihood of changes in the active growth site (Fig.~\ref{fig:fig2}a). We first pre-grow the precipitation pattern for approximately 100~s, then stop the cobalt injection and wait for a controlled interval $t_w=10$–$90$~s. After this wait time, the pump is restarted briefly and the locations at which growth resumes are recorded. These sites may coincide with previously active regions or emerge at new positions. Representative before-and-after images of this breach-induced regrowth are shown in Fig.~\ref{fig:fig2}b.

The age of the precipitate at each renewed growth site is obtained from heat maps such as those in Fig.~\ref{fig:fig2}c, which record the time $t$ at which each point last precipitated relative to the stopping time $t=0$. Small $t$ corresponds to freshly formed material, whereas large $t$ indicates older regions. The relevant age of a site at restart is therefore $t+t_w$.

Before analyzing the dependence on breach age and wait time, we first verify that the underlying distribution of precipitate ages is unbiased. To this end, we extract the formation times $t_{\mathrm{edge}}$ along the pattern boundary and plot their ordered values versus rank (not shown). In all experiments, this relation is linear, indicating a uniform distribution of formation times along the edge. The slope varies slightly with silicate concentration $c$, consistent with changes in filament width and overall pattern compactness. Linear behavior is also obtained when the entire interior of the pattern is included in the analysis. 

\begin{figure}[t]
\centering
\includegraphics[width=0.96\columnwidth]{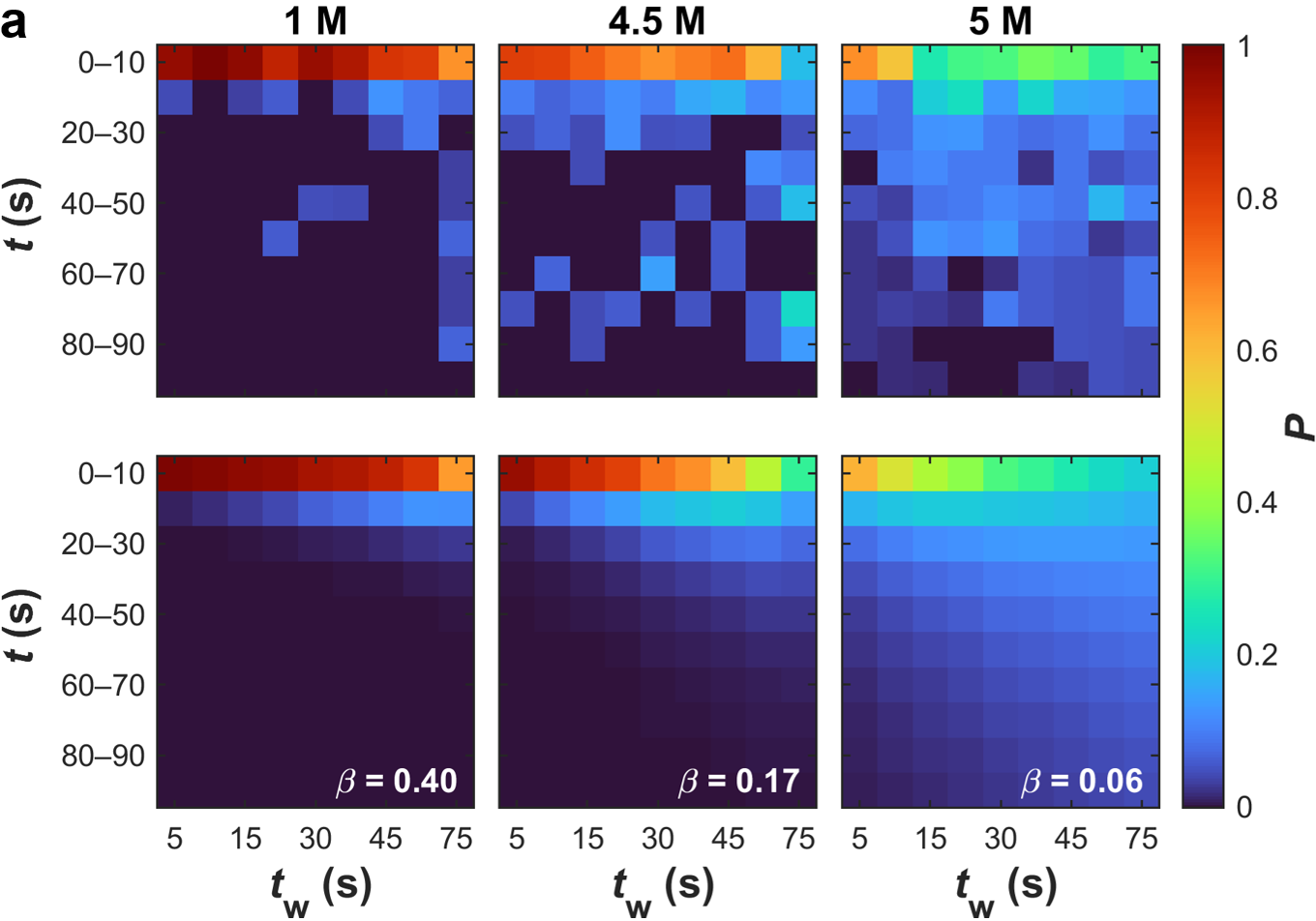}\\[6pt]
\includegraphics[width=0.99\columnwidth]{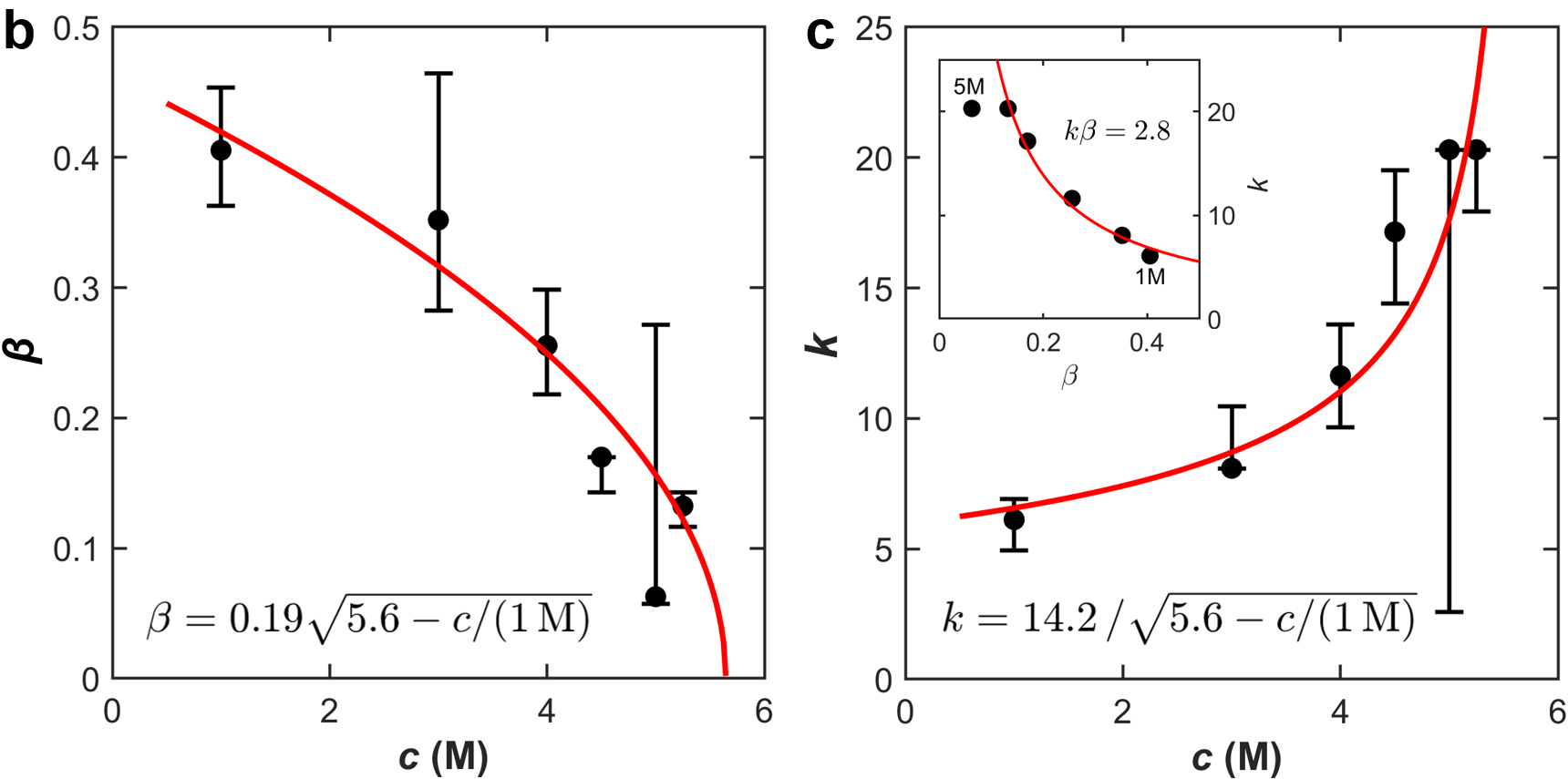}
\caption{(a) Age-dependent breach probabilities after pauses in $\mathrm{CoCl_2}$ injection at varying silicate concentrations $c$ between 1~M and 5~M. Top row: experimental probability distributions $P(t|t_w)$ per 10-s bin for different breach formation times $t$ and wait times $t_w$. Each column is individually normalized. Bottom row: stretched-exponential fits with respective $\beta$ values. 
(b,c) Stretched exponential fit parameters $\beta$ and $k$ versus $c$. Error bars represent profile-likelihood-based confidence intervals. Fits (red curves) yield the indicated square root laws. Inset: $k$ versus $\beta$ follows $k\,\beta$~=~2.8. The 5~M data point was excluded from fits due to poorly constrained values (flat profile likelihood).
}
\label{fig:analysis}
\end{figure}

The top row of Fig.~\ref{fig:analysis}a shows the distribution of breach events as a function of the wait time $t_w$ and precipitate age $t$. The data comprise 270 experiments (five per $(c,t_w)$ pair) with an average of $4.8\pm1.6$ detected breach sites per run. This number is independent of $t_w$ and shows only a mild decrease from 4.8 to 3.9 as $c$ increases from 1.0 to 4.5~M, with a subsequent increase at higher concentrations. Because the number of events fluctuates between runs, each column is normalized to unity, converting counts to probabilities while preserving the relative distribution across ages. Short wait times strongly favor breaches in the youngest regions of the membrane, consistent with the persistence of active growth zones in unperturbed systems. Increasing $t_w$ shifts breaches toward progressively older material, effectively sampling deeper into the membrane’s age distribution. This shift strengthens with increasing silicate concentration, concentrating probability in the lower-right region of the heat maps. 

The data in Fig.~\ref{fig:analysis}a allow us to test candidate functional forms for the aging kinetics. Because aging depends equally on the elapsed growth time $t$ and wait time $t_w$, the breach probability should depend only on their sum, $t+t_w$. In our previous work, we assumed a simple exponential bias, $P \sim e^{-kt}$, which here implies $P \sim e^{-k(t+t_w)}$. For fixed $t_w$, the prefactor $e^{-k t_w}$ factors out and is removed by column normalization, predicting identical distributions for all $t_w$. This prediction is inconsistent with the experiments, which show pronounced, concentration-dependent differences between columns. We therefore conclude that precipitate aging does not follow simple first-order kinetics.

We next compare alternative functional forms for the breach probability. Power laws, $P \sim (t+t_w)^{-\alpha}$, and stretched exponentials, $P \sim e^{-k(t+t_w)^{\beta}}$, both provide reasonable descriptions of the data. However, the power-law fits result in unphysically large exponents ($\alpha\approx 8$ for $c=$~1-4~M) and exhibit larger residual errors. In contrast, the stretched exponentials reproduce the measured probabilities more closely (Fig.~\ref{fig:analysis}a, lower row), yielding mean residual errors approximately 12\% lower than the power-law fits. Stretched-exponential kinetics are also widely observed in aging and relaxation processes in disordered materials such as glasses and gels, supporting their physical relevance in our system \cite{Sethna2022,Cipelletti2005,Amir2012,Vlad1996}.

Figures~\ref{fig:analysis}b,c summarize the fitted parameters $\beta$ and $k$ as functions of the silicate concentration $c$. The stretching exponent $\beta$ decreases from approximately 0.4 to 0.1 as $c$ increases from 1.0 to 5.25~M, indicating a progressively heavier-tailed distribution than a simple exponential. The data are well described by a square-root dependence, $\beta \propto (c^*-c)^{1/2}$, which extrapolates to a critical concentration $c^*=5.6$~M where $\beta \to 0$. The rate parameter $k$ increases with $c$ and follows the complementary scaling $k \propto (c^*-c)^{-1/2}$, diverging near the same $c^*$. Times are expressed in dimensionless units $t'/(1~\mathrm{s})$, so that $k$ is also dimensionless.

The inset of Fig.~\ref{fig:analysis}b shows $k$ versus $\beta$, revealing an inversely proportional dependence that follows $k\,\beta \approx 2.8$ and is consistent with the inverse and direct square-root laws of $k(c)$ and $\beta(c)$, respectively. The two parameters in the stretched-exponential probability distribution are therefore tightly coupled. This connection has important implications for the behavior of the material near the critical silicate concentration $c^*$. Since the exponent $\beta$ vanishes at $c^*$, one might initially interpret this state as a material without any age bias. However, the behavior is qualitatively different. Because $k$ diverges at the same point while the product $k\,\beta$ remains finite, the stretched exponential does not approach a bias-free, time-independent probability. Using the expansion $(t+t_w)^\beta \approx 1 + \beta \ln(t+t_w)$ for small $\beta$, we obtain
\begin{equation}
e^{-k (t+t_w)^{\beta}}
\approx
e^{-k}\,(t+t_w)^{-k\beta},
\label{eq:stretched_approx}
\end{equation}
so that the limiting form becomes a power law. The kinetics therefore lose a characteristic time scale and cross over to heavy-tailed, scale-free behavior, implying algebraically slow decay and long-lived memory. Increasing silicate concentration thus drives the membrane toward a dynamically critical state, consistent with the observed persistence of filaments. Comparable transient memory effects have been observed in cyclically driven non-Brownian suspensions \cite{Paulsen2014}. 

To gain further insight into the patterns formed at and above $c^*$, we performed experiments at concentrations of 5.5 and 6.0~M, close to the solubility limit of the sodium silicate used here. These conditions consistently produced filament-like patterns, indicating that the extrapolated critical concentration does not correspond to a transition toward radially uniform or age-independent growth. Instead, the persistence of localized growth zones is consistent with the power-law kinetics expected in the $\beta \to 0$ limit of the stretched-exponential model. Attempts to quantify the aging dynamics using our wait-time protocol were unsuccessful at these concentrations. Upon restarting injection, breaches occurred predominantly near the injection port rather than along the filament perimeter, suggesting that mechanical weakness dominated the failure process. These events likely represent experimental artifacts that obscure the intrinsic age selectivity of the membrane.

\begin{figure}[t]
\centering
\includegraphics[width=1.0\columnwidth]{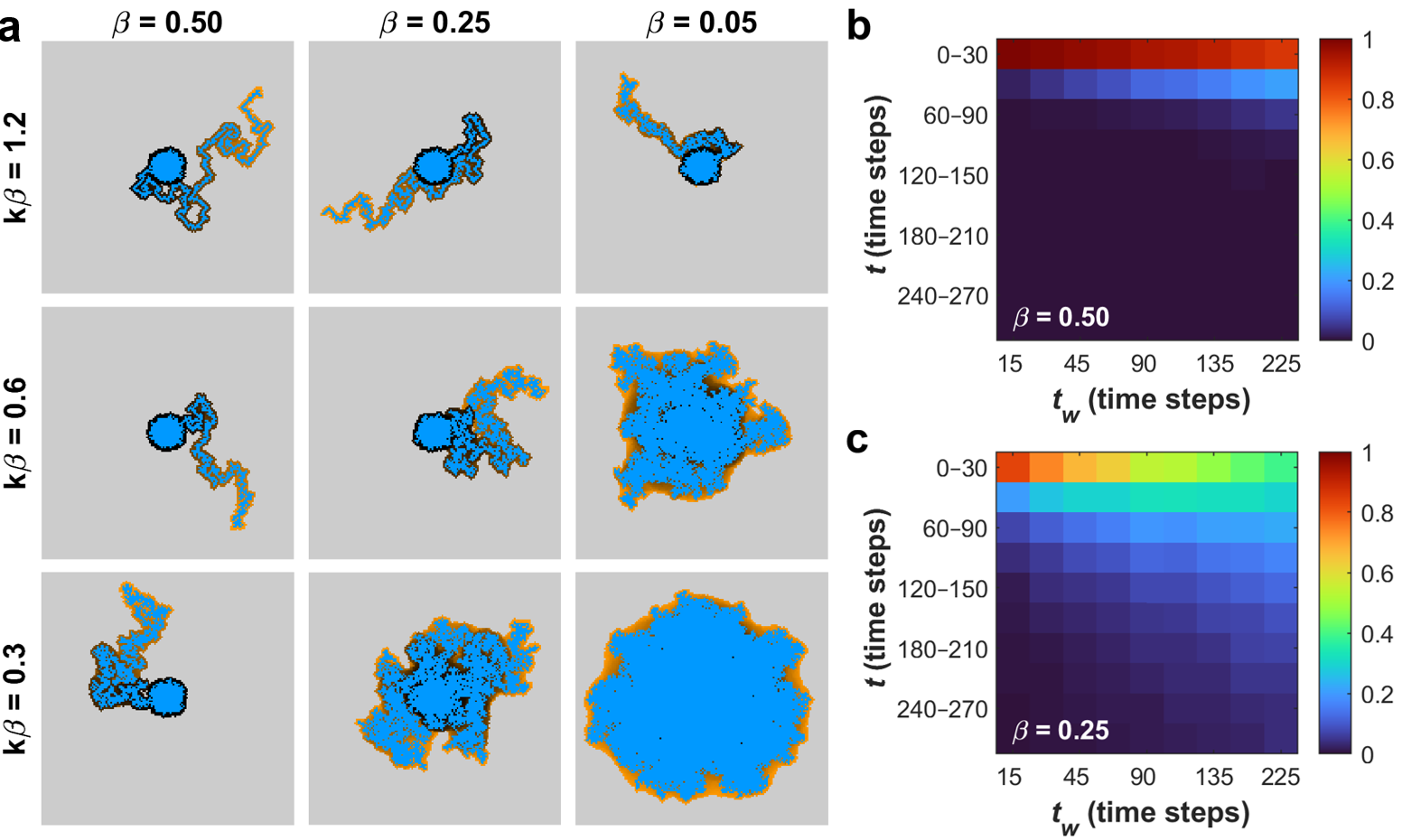}
\caption{
Stochastic cellular automaton simulations with an age-dependent breach probability $P(t)\propto \exp(-k t^{\beta})$.
(a) Representative morphologies for different $(\beta,k\,\beta)$ values showing filamentary conduits of increasing average width, irregular patterns, and rugged fronts. Blue represents the injected solution; orange and black mark newly formed and old precipitate, respectively. 
(b,c) Simulated restart experiments reproducing age/wait-time probability maps for $k\,\beta=1$ and $\beta = 0.5, 0.25$, respectively. Color denotes the column-normalized probability that the first breach  after a wait time $t_w$ occurs in material of age $t$.}
\label{fig:CA}
\end{figure}

Our finding of stretched exponentials governing the age-dependent dynamics of precipitate membranes can be further tested in a simple cellular automaton model \cite{Batista2023}. The model uses a two-dimensional lattice in which each site represents either reactant A (silicate), reactant B ($\mathrm{CoCl_2}$), or precipitate C. Precipitate forms locally whenever A and B coexist within a site’s neighborhood, and existing precipitate ages at each time step by incrementing its time stamp. Sites eligible for renewed growth are precipitate cells in contact with B; among these candidates, a single breach site is selected stochastically with probability weighted by an age-dependent kernel $P(t) \propto \exp(-kt^\beta)$, so that younger precipitate is preferentially removed. The displaced material is then re-deposited by converting the closest available A site to new precipitate, chosen randomly among the nearest candidates, which mimics incompressible flow that advances the membrane locally. 

Representative simulation results obtained with this model are shown in Fig.~\ref{fig:CA}a where rows and columns have the specified $k\,\beta$ and $\beta$ values, respectively. Note that the time scaling of the experiments does not transfer to the discrete time steps of the model, making a direct use of the measured $k\,\beta$ value impossible. Over the shown parameter range, we find primarily filaments but for small $k\,\beta$ values also irregular patterns and expanding fronts of approximate circular shape. The simulations also capture the experimentally observed thickening of filaments with decreasing values of $\beta$ (i.e., increasing silicate concentrations).

In addition, the model allows us to reproduce the wait-time procedure of our experiments (Fig.~\ref{fig:fig2}a). After a pre-growth phase of 1000 time steps, we pause the dynamics for a prescribed number of time steps $t_w$ by advancing all precipitate ages without permitting breaches or re-deposition, then restart the dynamics and record the age $t$ of the precipitate site selected for the first breach. Repeating this sampling generates model analogs of the experimental $t_w,t$ probability maps, with each column normalized to unity. Representative results for fixed $k\,\beta=1$ and two different $\beta$ are shown in Figs.~\ref{fig:CA}b,c. The simulations reproduce the experimentally observed broadening of the age distribution and the increasing $t_w$ dependence as $\beta$ decreases, confirming that the stretched-exponential aging kernel alone captures the persistence and thickening of filamentary growth.

While it is remarkable that such a simple model captures the essential experimental features, our simulations also reveal a limitation. For irregular patterns and expanding fronts, multiple growth zones remain active, whereas filamentary growth collapses to a single dominant site. This ``winner-take-all'' behavior is not observed experimentally. To break this behavior, we explored nonmonotonic aging functions that assign zero probability to sites with $t<t_{\mathrm{cut}}$. For the stretched exponential and tested parameters around $(k,\beta)=(4,0.25)$, this modification induces on average two active filaments when $t_{\mathrm{cut}}$ equals approximately 20 time steps (not shown), suggesting that a sharp suppression of breaches in the freshest precipitate may not have been resolved experimentally. However, alternative explanations include local curvature effects, local reactant depletion, or other self-limiting mechanisms such as the increasing pressure head in extending filament conduits.\\


In conclusion, we showed that precipitate membrane growth is governed by a universal aging law. Near a critical concentration $c^*$, the stretching exponent $\beta$ approaches zero, causing the system to lose its characteristic time scale and transition to scale-free, power-law kinetics. This behavior mirrors the "freezing" seen in glassy and jammed systems, where aging becomes logarithmically slow and failure is dictated by long-lived memory \cite{Cipelletti2005,Amir2012,Vlad1996,Berthier2011,Cugliandolo1993}. By identifying this transition, we establish age-controlled failure as a fundamental mechanism for pattern selection in far-from-equilibrium materials.

\begin{acknowledgments}
We thank Bruno C. Batista for initial help and advice.
\end{acknowledgments}

\section*{Data and Code Availability}
Experimental data and \textsc{MATLAB} (R2024b) scripts for analyses and cellular automaton simulations are available at https://github.com/osteinbock/Morris2026. Raw image data are available upon request.

\appendix

\section{Experimental Methods} 
Solutions of Na$_2$SiO$_3 \cdot$5H$_2$O (Fisher Chemicals) and CoCl$_2$ (TCL America) were prepared in high-purity water. The setup consisted of a horizontally placed LED monitor for homogeneous white or contrast-enhancing red illumination, the Hele-Shaw cell, a connected syringe pump (New Era Pump Systems, NE-4000), and a Nikon D3300 camera with a Tamron 90~mm macro lens mounted above the cell. The two Hele-Shaw plates (plexiglass) measured $21.5 \times 21.5$~cm$^2$ with a gap height of 0.2~mm created by small spacers. The central injection port faced downward. At the beginning of each experiment, 1~mL of sodium silicate solution was injected into the cell to produce an approximately circular layer around the port. During the experiments, images were recorded at a rate of 60~frames/s and spatial resolution of 35~pixel/mm. We constructed all age maps (e.g., Fig.~\ref{fig:fig2}c) based on the time of the largest local intensity drops. The injection rate was 0.15~mL/min, unless otherwise noted. All experiments were carried out at 21~$^\circ$C. 

\section{Numerical Methods} 
Most simulations were performed on a square lattice of $N \times N = 150 \times 150$ sites with an initial circular B region of radius $r_0 = 10$ lattice units. C sites are eligible for breach if they have at least $B_\mathrm{min} = 2$ neighboring B sites. At each time step, at most one breach occurs: probabilities are computed from the age-dependent kernel for all $m$ eligible sites, normalized to unity, and a single site is selected via multinomial sampling. Displaced material is redeposited at the nearest available A site, selected randomly among equidistant candidates. Simulations terminate when the outermost C site exceeds a radial distance of $N/2 - 2$ lattice units from the center.

For the wait-time analysis, simulations used $N = 250$ and first evolved for 1000 time steps to establish a mature pattern. Dynamics were then paused for $t_w$ steps during which all precipitate ages advance uniformly without permitting breaches or redeposition. Upon resumption, the age of the site selected for the first breach was recorded. For each $t_w$ value, this sampling was repeated 500 times; results were aggregated over 12 independent runs with different random seeds.


\begin{thebibliography}{99}

\bibitem{Sarkar2025}
S. Sarkar \textit{et al.}, 
Phys. Rev. Lett. \textbf{135}, 258301 (2025).

\bibitem{Li2022}
M.~Li \textit{et al.},
Nat.\ Rev.\ Mater.\ \textbf{7}, 235--249 (2022).

\bibitem{Shklyaev2024}
O.~E.~Shklyaev and A.~C.~Balazs,
Nat.\ Nanotechnol.\ \textbf{19}, 146--159 (2024),

\bibitem{Marchetti2013}
M.~C.~Marchetti \textit{et al.},
Rev.\ Mod.\ Phys.\ \textbf{85}, 1143--1189 (2013).

\bibitem{Knoll2024}
P.~Knoll, B.~Ouyang, and O.~Steinbock,
ACS Phys.\ Chem.\ Au \textbf{4}, 19--30 (2024).

\bibitem{Zhu2024}
W.~Zhu, P.~Knoll, and O.~Steinbock,
J.\ Phys.\ Chem.\ Lett.\ \textbf{15}, 5476--5487 (2024),

\bibitem{Keim2014}
N.~C.~Keim and P.~E.~Arratia,
Phys.\ Rev.\ Lett.\ \textbf{112}, 028302 (2014),

\bibitem{Paulsen2014}
J.~D.~Paulsen, N.~C.~Keim, and S.~R.~Nagel,
Phys.\ Rev.\ Lett.\ \textbf{113}, 068301 (2014).

\bibitem{Cipelletti2005}
L.~Cipelletti and L.~Ramos,
J.\ Phys.: Condens.\ Matter \textbf{17}, R253 (2005).

\bibitem{AguilarHidalgo2018}
D.~Aguilar-Hidalgo, S.~Werner, O.~Wartlick, M.~González-Gaitán, B.~M.~Friedrich, and F.~Jülicher,
Phys.\ Rev.\ Lett.\ \textbf{120}, 198102 (2018).

\bibitem{Trepat2009}
X.~Trepat, M.~R.~Wasserman, T.~E.~Angelini, E.~Millet, D.~A.~Weitz, J.~P.~Butler, and J.~J.~Fredberg,
Nat.\ Phys.\ \textbf{5}, 426--430 (2009).

\bibitem{Dervaux2008}
J.~Dervaux and M.~Ben Amar,
Phys.\ Rev.\ Lett.\ \textbf{101}, 068101 (2008).

\bibitem{Barge2015}
L.~M.~Barge \textit{et al.},
Chem.~Rev. \textbf{115}, 8652--8703 (2015).

\bibitem{Nakouzi2016}
E.~Nakouzi and O.~Steinbock,
Sci.\ Adv.\ \textbf{2}, e1601144 (2016).

\bibitem{Cardoso2020}
S.~S.~S.~Cardoso, J.~H.~E.~Cartwright, J.~Čejková, L.~Cronin, A.~De~Wit, S.~Giannerini, D.~Horváth, A.~Rodrigues, M.~J.~Russell, C.~I.~Sainz-Díaz, and Á.~Tóth,
Artif.\ Life \textbf{26}, 315--326 (2020).

\bibitem{Pantaleone2009}
J.~Pantaleone, Á.~Tóth, D.~Horváth, L.~RoseFigura, W.~Morgan, and J.~Maselko,
Phys.\ Rev.\ E \textbf{79}, 056221 (2009).

\bibitem{Cartwright2002}
J.~H.~E.~Cartwright, J.~M.~Garc\'{\i}a-Ruiz, M.~L.~Novella, and F.~Ot\'alora,
J.\ Colloid Interface Sci.\ \textbf{256}, 351--359 (2002).

\bibitem{Guler2023}
B.~Aslanbay~G\"uler, Z.~Demirel, and E.~Imamoglu,
Langmuir \textbf{39}, 13611--13619 (2023).

\bibitem{Bohner2015}
B.~Bohner, G.~Schuszter, D.~Horváth, and Á.~Tóth,
Chem.\ Phys.\ Lett.\ \textbf{631--632}, 114--117 (2015).

\bibitem{Rauscher2018}
E.~Rauscher, G.~Schuszter, B.~Bohner, Á.~Tóth, and D.~Horváth,
Phys.\ Chem.\ Chem.\ Phys.\ \textbf{20}, 5766--5770 (2018).

\bibitem{Zheng2026}
M.~Zheng, E.~Dumont, R.~D.~Featherstone, N.~Mackay, H.~E.~Huppert, J.~H.~E.~Cartwright, and A.~F.~Routh,
Soft Matter - available online (2026).

\bibitem{Zheng2025}
M.~Zheng, P.~R.~L.~Welche, S.~S.~S.~Cardoso, H.~E.~Huppert, J.~H.~E.~Cartwright, and A.~F.~Routh,
Philos.\ Trans.\ A \textbf{383}, 20240266 (2025).

\bibitem{ThouvenelRomans2003}
S.~Thouvenel-Romans and O.~Steinbock,
J.\ Am.\ Chem.\ Soc.\ \textbf{125}, 4338--4341 (2003).

\bibitem{Batista2014}
B.~C.~Batista, P.~Cruz, and O.~Steinbock,
Langmuir \textbf{30}, 9123--9129 (2014).

\bibitem{Kubodera2025}
Y.~Kubodera, M.~Matsuo, and S.~Nakata,
Phys.\ Chem.\ Chem.\ Phys.\ \textbf{27}, 18454--18458 (2025).

\bibitem{ThouvenelRomans2004}
S.~Thouvenel-Romans, W.~van Saarloos, and O.~Steinbock,
Europhys.\ Lett.\ \textbf{67}, 42 (2004).

\bibitem{Nogueira2023}
J.~A.~Nogueira, B.~C.~Batista, M.~A.~Cooper, and O.~Steinbock,
Angew.\ Chem.\ Int.\ Ed.\ \textbf{62}, e202306885 (2023).

\bibitem{Haudin2014}
F.~Haudin, J.~H.~E.~Cartwright, F.~Brau, and A.~De~Wit,
Proc.~Natl.~Acad.~Sci.~U.S.A. \textbf{111}, 17363--17367 (2014).

\bibitem{Wagatsuma2017}
S.~Wagatsuma, T.~Higashi, Y.~Sumino, and A.~Achiwa,
Phys.~Rev.~E \textbf{95}, 052220 (2017).

\bibitem{Rocha2021}
L.~A.~M.~Rocha, J.~H.~E.~Cartwright, and S.~S.~S.~Cardoso,
Phys.~Chem.~Chem.~Phys. \textbf{23}, 5222--5235 (2021).

\bibitem{Rocha2022}
L.~A.~M.~Rocha, L.~Thorne, J.~J.~Wong, J.~H.~E.~Cartwright, and S.~S.~S.~Cardoso,
Langmuir \textbf{38}, 6700--6710 (2022).

\bibitem{Rieder2022}
J.~Rieder, L.~Nicoleau, F.~Glaab, A.~E.~S.~Van~Driessche, 
J.~M.~Garcia-Ruiz, W.~Kunz, and M.~Kellermeier,
J.\ Colloid Interface Sci.\ \textbf{618}, 206--218 (2022).

\bibitem{Facchini2025}
G.~Facchini, M.~A.~Budroni, G.~Schuszter, F.~Brau, and A.~De~Wit,
Phys.\ Rev.\ Lett.\ \textbf{135}, 018001 (2025).

\bibitem{Batista2023}
B.~C.~Batista, A.~Z.~Morris, and O.~Steinbock,
Proc.~Natl.~Acad.~Sci.~U.S.A. \textbf{120}, e2305172120 (2023).

\bibitem{Zahoran2019}
R.~Zahor\'an, \'A.~Kukovecz, \'A.~T\'oth, D.~Horv\'ath, and G.~Schuszter,
Phys.~Chem.~Chem.~Phys. \textbf{21}, 11345--11350 (2019).

\bibitem{Wang2017}
Q.~Wang, M.~R.~Bentley, and O.~Steinbock,
J.~Phys.~Chem.~C \textbf{121}, 14120--14127 (2017).

\bibitem{Cartwright2011}
J.~H.~E.~Cartwright, B.~Escribano, and C.~I.~Sainz-Díaz,
Langmuir \textbf{27}, 3286--3293 (2011).

\bibitem{Haudin2015}
F.~Haudin, V.~Brasiliense, J.~H.~E.~Cartwright, F.~Brau, and A.~De~Wit,
Phys.\ Chem.\ Chem.\ Phys.\ \textbf{17}, 12804--12811 (2015).

\bibitem{Haudin2015b}
F.~Haudin, J.~H.~E.~Cartwright, and A.~De~Wit,
J.\ Phys.\ Chem.\ C \textbf{119}, 15067--15076 (2015).

\bibitem{Sethna2022}
J.~P.~Sethna,
Nat.\ Rev.\ Phys.\ \textbf{4}, 501--503 (2022).

\bibitem{Amir2012}
A.~Amir, Y.~Oreg, and Y.~Imry,
Proc.\ Natl.\ Acad.\ Sci.\ U.S.A.\ \textbf{109}, 1850--1855 (2012).

\bibitem{Vlad1996}
M.~O.~Vlad, R.~Metzler, T.~F.~Nonnenmacher, and M.~C.~Mackey,
J.\ Math.\ Phys.\ \textbf{37}, 2279--2306 (1996).

\bibitem{Berthier2011}
L.~Berthier and G.~Biroli,
Rev.\ Mod.\ Phys.\ \textbf{83}, 587--645 (2011).

\bibitem{Cugliandolo1993}
L.~F.~Cugliandolo and J.~Kurchan,
Phys.\ Rev.\ Lett.\ \textbf{71}, 173--176 (1993).

\end{thebibliography}
\end{document}